\documentclass[seceq]{ptptex}
\usepackage{wrapft}

\newcommand{\btem}{\bibitem}
\newcommand{\beq}{\begin{eqnarray}}
\newcommand{\eeq}{\end{eqnarray}}

\renewcommand{\d}{\partial}

\def\Vec#1{\mbox{\boldmath$#1$}}
\def\Ren#1{\mbox{\boldmath$#1$}}

\title{%
New forms of non-relativistic
and relativistic hydrodynamic
 equations as derived by the renormalization-group method
}

\subtitle{possible functional  ansatz in the moment method
consistent with Chapman-Enskog theory}    

\author{
  Kyosuke \textsc{Tsumura}$^{1,}$\footnote{E-mail: kyousuke\_tsumura@fujifilm.co.jp}
  and
  Teiji \textsc{Kunihiro}$^{2,}$\footnote{E-mail: kunihiro@ruby.scphys.kyoto-u.ac.jp}
}

\inst{
  ${}^1$Analysis Technology Center,
  Fujifilm Corporation,
  Kanagawa 250-0193, Japan
  \\
  ${}^2$Department of Physics,
  Kyoto University,
  Kyoto 606-8502, Japan
}

\abst{
After a brief account of the derivation of 
the first-order relativistic hydrodynamic
equation as a construction of the 
invariant manifold of relativistic Boltzmann equation,
we give a sketch of derivation of the second-order hydrodynamic equation (extended
thermodynamics) both in the nonrelativistic and relativistic cases. 
We show that the resultant equation suggests 
a novel ansatz for the functional form to be used in 
Grad moment method, which turns out to give the same
expressions for the transport coefficients 
as those given in the Chapman-Enskog theory as well as the
novel expressions for the relaxation times and lenghts allowing natural physical 
interpretaion.
}

\begin{document}

\maketitle

\section{Introduction}
  The relativistic hydrodynamic equation is widely and successfully used
  in nuclear and astrophysics:
  The dynamical evolution of
  the hot and/or dense QCD matter
  produced in the Relativistic Heavy Ion Collider (RHIC) experiments
  can be well described
  by relativistic hydrodynamic simulations \cite{qcd001,qcd003}.
  The suggestion that the created matter may have only a tiny viscosity
  prompted an interest in the origin of the viscosity in the
  created matter to be described using the relativistic quantum field theory
  and also the dissipative hydrodynamic equations.
  The relativistic dissipative hydrodynamic equation
  has been also applied to various high-energy astrophysical phenomena\cite{ast001},
  e.g.,
  the accelerated expansion of the universe caused by the bulk viscosity of dark
  matter and/or dark energy \cite{ast002}.
  We must, however, ask if 
  the theory of relativistic hydrodynamics for viscous fluids is fully established?
  The answer is, unfortunately, no.
  
  The fundamental problems
  on the  relativistic dissipative hydrodynamic equation may be summarized
  as follows:
  (a)~ambiguities  in the definition of the flow velocity
  \cite{hen001,hen002}
  , (b)~unphysical instabilities of the equilibrium state \cite{hyd002},
  and
  (c)~lack of causality \cite{mic001,hen003,mic004}.

  Here, we note that the non-equilibrium process may evolve
  through  some stages of hierarchical  dynamics\cite{text}:
  In the beginning of the time-evolution of an isolated prepared state,
  the whole dynamical evolution of the system
  will be governed by Hamiltonian dynamics
  that is time-reversal invariant.
  As  time goes on,
  the dynamics is relaxed into the kinetic regime
  where the time-evolution of the system is well described by
   kinetic equations which describes a coarse-grained slower dynamics:
 The Boltzmann equation for the one-body distribution function
  is one of them\cite{text};
 Usually the original time-reversal invariance is lost in the description 
by the kinetic equation through the coarse-graining.
  As the system is further relaxed,
  the time evolution will be described in terms of the hydrodynamic quantities,
  i.e.,  the flow velocity,   particle-number density,
  and local temperature.
  In this sense,
  the hydrodynamics is the far-infrared asymptotic dynamics of
  the kinetic equation.

  In this report, we take
 the relativistic Boltzmann equation (RBE)\cite{mic001} as a typical kinetic equation
 and explore the basic problems with the relativistic hydrodynamics.
For obtaining the proper relativistic hydrodynamic equation, 
it is a legitimate and natural way to start with 
RBE which is Lorentz invariant and does not have  stability nor causality problems\cite{mic001}.
For analyzing the problems (a) and (b) first,
we derive hydrodynamic equations\cite{env009,Tsumura:2011cj} for a viscous fluid from RBE
on the basis of a mechanical reduction theory called the renormalization-group(RG) 
method\cite{rgm001,env001,env006}
 and a natural ansatz on the origin of dissipation.
We then proceed to the causality problem by deriving the so called
 extended thermodynamics\cite{extended,meso,Gorban}: We shall derive the mesoscopic
dynamics of the RBE by constructing the invariant/attractive manifold incorporating
the first excited as well as the zero modes of the linearized collision operator.
We obtain the same expressions for the transport coefficients as given by the
Chapman-Enskog method\cite{chapman} and also the relaxation times in terms of relaxation functions
which have natural physical interpretation. 
We also show that the resultant equation suggests 
a novel ansatz for the functional form to be used in 
the  moment method\cite{grad}.

  \setcounter{equation}{0}
\section{Relativistic Boltzmann equation}

The relativistic Boltzmann equation  reads\cite{mic001}
  \begin{eqnarray}
    \label{eq:1-001}
    p^\mu \, \partial_\mu f_p(x) = C[f]_p(x),
  \end{eqnarray}
where $f_p(x)$ denotes the one-particle distribution function
  defined in the phase space $(x \,,\, p)$;
$p^{\mu}$ denotes  the four-momentum of the on-shell particle, i.e.,
 $p^\mu p_\mu = p^2 = m^2$ and $p^0 > 0$.
 The r.h.s. is the collision integral 
  \begin{eqnarray}
    \label{eq:1-002}
    C[f]_p(x) \equiv \frac{1}{2!}\sum_{p_1} \, \frac{1}{p_1^0}
    \sum_{p_2} \frac{1}{p_2^0} \sum_{p_3}  \frac{1}{p_3^0} 
    \omega(p, p_1|p_2, p_3)\Big( f_{p_2}(x) f_{p_3}(x) - f_p(x) f_{p_1}(x) \Big),
  \end{eqnarray}
  where $\omega(p \,,\, p_1|p_2 \,,\, p_3)$ denotes
  the transition probability due to the microscopic two-particle interaction.
  Owing to the symmetry property
$\omega(p, p_1|p_2, p_3) = \omega(p_2, p_3|p, p_1)$
$=\omega(p_1, p|p_3, p_2) = \omega(p_3 , p_2|p_1 , p)$
and the energy-momentum conservation,
the collision operator satisfies the following identity
for an arbitrary vector $\varphi_p(x)$,
  \begin{eqnarray}
    \label{eq:coll-symm}
    \sum_p \, \frac{1}{p^0} \, \varphi_p(x) \, C[f]_p(x)
    &=& \frac{1}{2!} \,\frac{1}{4}\, \sum_{p} \, \frac{1}{p^0}\,
    \sum_{p_1} \, \frac{1}{p^0_1}\,
    \sum_{p_2} \, \frac{1}{p^0_2}\,
    \sum_{p_3} \, \frac{1}{p^0_3}\,
    \nonumber\\
    &&
    \times
    \omega(p \,,\, p_1|p_2 \,,\, p_3)\,
    \Big(\varphi_p(x) \,+\,\varphi_{p_1}(x)\,- \,\varphi_{p_2}(x)\, -\,\varphi_{p_3}(x)\Big)
    \nonumber \\
    &&
    \times
    \Big( f_{p_2}(x) \, f_{p_3}(x) - f_p(x) \, f_{p_1}(x) \Big),
  \end{eqnarray}
from which one can show that
the function $\varphi_p(x)=a(x) + p^{\mu} \, b_{\mu}(x)$ 
is a collision invariant satisfying the equation
$ \sum_p \, \frac{1}{p^0} \, \varphi_p(x) \, C[f]_p(x) = 0$,
where $a(x)$ and $b_{\mu}(x)$ being arbitrary functions of $x$.
 This form is, in fact, the most general form of a collision invariant \cite{mic001}.

  \setcounter{equation}{0}
  \section{Reduction to hydrodynamic equation}

  Before developing our analysis based on the RG method,
  we briefly summarize  ad hoc aspects in the standard methods
  such as the Chapman-Enskog expansion and Maxwell-Grad moment methods
 \cite{mic001}.
  
  In  Chapman-Enskog  method\cite{chapman},
 the zeroth-order solution is given by the local equilibrium 
  distribution function, i.e., the J\"{u}ttner function\cite{juettner}
$ f^{(0)}_p(x) = f^{\mathrm{eq}}_p(x)$.
Then, one imposes a {\em condition of fit} or {\em matching condition}, 
which usually consists of  drastic assumptions that
  internal energy and the particle-number density
  in the {\it nonequilibrium} state are the same as those
  in the local {\it equilibrium} state\cite{mic001}.
  Indeed these constraints  are equivalent to
  a  physical assumption that
  there is no internal energy nor particle-number density of  dissipative origin.
 Such an assumption, however,  does not have any solid physical foundation, 
  since the distribution function in the nonequilibrium state
 should be quite different from that in the local equilibrium state.
Such matching conditions are also imposed in the moment method\cite{mic001}.

  In the RG method that we adopt\cite{env001,env006},
  one needs no such matching conditions for derivation,
  and their correct forms are obtained as a property of the derived equation
  once the frame is specified by the macroscopic-frame vector\cite{env009,Tsumura:2011cj}.
  We have shown \cite{env009,Tsumura:2011cj} that
  the conditions of fit in the energy frame
  are compatible with the underlying Boltzmann equation,
  but those in the particle frame adopted in the literature are not
  and thus will never be satisfied in any physical system.

We  solve RBE (\ref{eq:1-001})
  in the hydrodynamic regime, and thereby derive the hydrodynamic equations 
governing the hydrodynamic variables which are introduced to parametrize the 
distribution function.
  To make a coarse graining with  Lorentz covariance\cite{env009,Tsumura:2011cj},
  we introduce a timelike Lorentz vector $\Ren{a}^\mu$,
  with $\Ren{a}^0\,>\,0$.  Thus, $\Ren{a}^\mu$  specifies
  the covariant and macroscopic coordinate system where
  the velocity field of the hydrodynamic flow is defined.
We call $\Ren{a}^\mu$ the \textit{macroscopic-frame vector}. 
  Although $\Ren{a}^\mu$ could depend on the momentum $p$ as well as the
  space-time coordinate $x$, i.e.,
$\Ren{a}^\mu = \Ren{a}^\mu_p(x)$,
the possible momentum dependence may not be legitimate as a macroscopic
frame vector\footnote{As is shown in \citen{env009,Tsumura:2011cj},
the Eckart frame can be realized only when the macroscopic frame vector
has a $p$ dependence. This implies that the macroscopic space-time
in the Eckart frame
is defined
for respective particle state with a definite energy-momentum, which may 
have a difficulty in a physical interpretation\cite{next}. In fact,
it can be shown\cite{next} that 
the $p$-independent timelike macroscopic frame vector
with the Lorentz covariance uniquely leads to the hydrodynamics
in the Landau-Lifshitz frame.
}.
Once $\Ren{a}^\mu$ is given,  RBE(\ref{eq:1-001}) 
is rewritten in terms of the new coordinates $(\tau \,,\, \sigma^\mu)$ as
  \begin{eqnarray}
    \label{eq:start}
    \frac{\partial}{\partial \tau} f_p(\tau \,,\, \sigma)
    = \frac{1}{p \cdot \Ren{a}_p(\tau \,,\, \sigma)} \, C[f]_p(\tau \,,\, \sigma)
    - \varepsilon \, \frac{1}{p \cdot \Ren{a}_p(\tau \,,\, \sigma)}
    \, p \cdot \Ren{\nabla} f_p(\tau \,,\, \sigma),
  \end{eqnarray}
where
$
{\partial}/{\partial\tau} \equiv (1/{\Ren{a}^2_p(x)})\, \Ren{a}^\nu_p(x)\,\partial_\nu$ and
$  \Ren{\nabla}^\mu \equiv \Ren{\Delta}^{\mu\nu}_p(x)\,\partial_\nu\equiv  {\d}/{\d \sigma_{\mu}}$,
with 
$\Ren{\Delta}^{\mu\nu}_p(x) \equiv  g^{\mu\nu} - {\Ren{a}_p^\mu(x) \, \Ren{a}_p^\nu(x)}/{\Ren{a}_p^2(x)}$.
In Eq.(\ref{eq:start}),
the small parameter $\varepsilon$ represents  the space-nonuniformity, which 
  may be identified with  the ratio of the average particle distance over the mean free path,
  i.e., the Knudsen number;
$\varepsilon$ will be set back to unity in the final stage.
This seemingly trivial rewrite of the equation  expresses the assumption
  that only the spatial inhomogeneity is the origin of the dissipation.
 It is noteworthy that
  the RG method applied to the nonrelativistic Boltzmann equation
  with the corresponding assumption successfully leads 
  to the Navier-Stokes equation \cite{env007};
  the present approach\cite{env009,Tsumura:2011cj}
 is simply a covariantization of the nonrelativistic case.

In the RG method \cite{env001,env006,env009,Tsumura:2011cj}, 
we first try to obtain the perturbative solution $\tilde{f}_p$ 
  to Eq. (\ref{eq:start})
  around the arbitrary initial time $\tau = \tau_0$
  with the initial value $f_p(\tau_0 ,\, \sigma)$;
   $ \tilde{f}_p(\tau = \tau_0 \,,\, \sigma \,;\, \tau_0) = f_p(\tau_0 \,,\, \sigma)$.
Here we have made explicit that the solution depends on the initial time $\tau_0$;
  we suppose that the initial value is on an exact solution.
  The initial value as well as the solution is expanded
  with respect to $\varepsilon$ as follows:
$\tilde{f}_p(\tau, \sigma ; \tau_0)
    = \tilde{f}_p^{(0)}(\tau , \sigma ; \tau_0)
    + \varepsilon  \tilde{f}_p^{(1)}(\tau ,\, \sigma ; \tau_0)
    + \cdots$, and an obviously similar form for $f_{p}(\tau_0, \sigma)$. 
    
It is easy to see that the zeroth-order equation reads
  \begin{eqnarray}
    \label{eq:1-025}
    \frac{\partial}{\partial \tau} \tilde{f}^{(0)}_p(\tau \,,\, \sigma \,;\, \tau_0)
    = \frac{1}{p \cdot \Ren{a}_p(\sigma \,;\, \tau_0)} \,
    C[\tilde{f}^{(0)}]_p(\tau \,,\, \sigma \,;\, \tau_0).
  \end{eqnarray}
  Since we are interested in the slow motion
  that would be realized asymptotically as $\tau \rightarrow \infty$,
  we should take the following stationary solution or the fixed point,
   $\frac{\partial}{\partial \tau}\tilde{f}_p^{(0)}(\tau \,,\, \sigma \,;\, \tau_0) = 0$,
  which is realized when
 $C[\tilde{f}^{(0)}]_p(\tau \,,\, \sigma \,;\, \tau_0) = 0$,
  for arbitrary $\sigma$.
  Thus, 
we find that the zero-th order solution $\tilde{f}_p^{(0)}(\tau \,,\, \sigma \,;\, \tau_0)$
  is  a local equilibrium distribution function
   \begin{eqnarray}
    \label{eq:1-028}
    \tilde{f}_p^{(0)}(\tau \,,\, \sigma \,;\, \tau_0)
    = \frac{1}{(2\pi)^{3}} \,
    \exp \Bigg[ \frac{\mu(\sigma \,;\, \tau_0)
        - p^\mu \, u_\mu(\sigma \,;\, \tau_0)}{T(\sigma \,;\, \tau_0)} \Bigg]
    \equiv f^{\mathrm{eq}}_p(\sigma \,;\, \tau_0),
  \end{eqnarray}
  with $u^\mu(\sigma \,;\, \tau_0) \, u_\mu(\sigma \,;\, \tau_0) = 1$.
  Note that the  would-be integration constants
  $T(\sigma \,;\, \tau_0)$, $\mu(\sigma \,;\, \tau_0)$, and $u_\mu(\sigma \,;\, \tau_0)$
  are independent of $\tau$ but may depend on $\tau_0$ as well as $\sigma$.
  
Given the zero-th order solution, the first-order equation reads
  \begin{eqnarray}
    \label{eq:1-030}
    \frac{\partial}{\partial \tau} \tilde{f}_p^{(1)}(\tau)
    = \sum_q \, A_{pq} \, \tilde{f}_q^{(1)}(\tau) 
+ F_p,
 \end{eqnarray}
where $A_{pq}$ denotes the matrix element of the linearized collision operator 
$A$ given by
  \begin{eqnarray}
    \label{eq:1-031}
    (A)_{pq}=A_{pq} \equiv \frac{1}{p \cdot \Ren{a}_p} \,
    \frac{\partial}{\partial f_q} C[f]_p \, \Bigg|_{f =  f^{\mathrm{eq}}}.
  \end{eqnarray}
and
\beq
 F_p\equiv - \frac{1}{p \cdot \Ren{a}_p} \, p \cdot \Ren{\nabla} f^{\mathrm{eq}}_p.
\eeq
 It is essential for obtaining a slow motion
to clarify the spectral properties of $A$.
 For this purpose, it is found convenient
to convert $A$ to another linear operator,
\beq
 L \equiv (f^{\mathrm{eq}})^{-1} \, A \, f^{\mathrm{eq}},
\eeq
  with the diagonal matrix
  $f^\mathrm{eq}_{pq} \equiv f^\mathrm{eq}_p \, \delta_{pq}$.
To characterize the properties of the linear operator $L$,
let us define an inner product
  between arbitrary nonzero vectors $\varphi$ and $\psi$ by
  \begin{eqnarray}
    \label{eq:1-034}
    \langle  \, \varphi \,,\, \psi \, \rangle
    \equiv \sum_{p} \, \frac{1}{p^0} \, (p \cdot \Ren{a}_p) \,
    f^{\mathrm{eq}}_p \, \varphi_p \, \psi_p.
  \end{eqnarray}
Then it can be shown\cite{env009,Tsumura:2011cj} 
that the linearized collision operator has a remarkable properties that 
it is semi-negative definite and has five zero modes  given by
  \begin{eqnarray}
    \label{eq:1-039}
    \varphi_{0p}^\alpha \equiv \left\{
    \begin{array}{ll}
      \displaystyle{p^\mu} & \displaystyle{\mathrm{for}\,\,\,\alpha = \mu}, \\[2mm]
      \displaystyle{1\times m}     & \displaystyle{\mathrm{for}\,\,\,\alpha = 4}.
    \end{array}
    \right.
  \end{eqnarray}
The functional subspace spanned by the five zero modes is called P$_0$ space and
the projection operator to it is denoted by $P_0$; we define $Q_0$
by $Q_0=1-P_0$. In the following, we also use the modified projection operators 
defined by
$\bar{P}_0=f^{\rm eq}P_0(f^{\rm eq})^{-1}$ and 
$\bar{Q}_0=f^{\rm eq}Q_0(f^{\rm eq})^{-1}$.

We proceed to solve the perturbed equations up to the second order.
Then summing up the perturbative solutions,
  we have an approximate solution around $\tau \simeq \tau_0$:
$\tilde{f}_p(\tau \,,\, \sigma \,;\, \tau_0)
    = \tilde{f}^{(0)}_p(\tau \,,\, \sigma \,;\, \tau_0)
    + \varepsilon \, \tilde{f}^{(1)}_p(\tau \,,\, \sigma \,;\, \tau_0)
    + \varepsilon^2 \, \tilde{f}^{(2)}_p(\tau \,,\, \sigma \,;\, \tau_0)
    + O(\varepsilon^3)$.
  It should be noted  that this solution contains
  the secular terms that apparently invalidates the perturbative expansion
  for $\tau$ away from the initial time $\tau_0$.
  We can, however, utilize the secular terms to obtain
  an asymptotic solution valid in a global domain\cite{env001,env006}.
  Indeed   we have a family of curves
  $\tilde{f}_p(\tau \,,\, \sigma \,;\, \tau_0)$
  parameterized with $\tau_0$:
  They are all on the exact solution
  $f_p(\sigma \,;\, \tau)$ at $\tau = \tau_0$ up to $O(\varepsilon^3)$,
  although only valid  for $\tau$ near $\tau_0$ locally.
  Then,  the {\em envelope curve} of the family of curves,
  which is in contact with each local solution at $\tau = \tau_0$, will
  give a global solution in our asymptotic situation, which is shown to 
be the case\cite{env001,env006}.
  According to the classical theory of envelopes,
  the envelope that is in contact with any curve in the family
  at $\tau = \tau_0$ is obtained\cite{env001} by
  \begin{eqnarray}
    \label{eq:1-058}
    \frac{d}{d\tau_0}
    \tilde{f}_p(\tau \,,\, \sigma \,;\, \tau_0) \Bigg|_{\tau_0 = \tau} = 0.
  \end{eqnarray}
The derivative  w.r.t. $\tau_0$ hits the hydrodynamic variables, and hence we have 
the evolution equation of them that is identified with the hydrodynamic 
equation\cite{env007,env009}. We also note that the invariant manifold 
which corresponds to the hydrodynamics
 in the functional space of the distribution function is explicitly obtained in the
present method\cite{env009,Tsumura:2011cj}.
  
  Putting back $\varepsilon = 1$,
Eq.(\ref{eq:1-058}) is nicely reduced to the following form,
  \begin{eqnarray}
    \label{eq:1-061}
    \sum_{p} \, \frac{1}{p^0} \, \varphi_{0p}^\alpha \,
    \Bigg[ (p \cdot \Ren{a}_p) \, \frac{\partial}{\partial \tau}
      + \varepsilon \, p \cdot \Ren{\nabla} \Bigg]
    \Big\{ f^{\mathrm{eq}}_p
    -  \, \big[ A^{-1} \, \bar{Q}_0 \, F \big]_p \Big\} + O(\varepsilon^3) = 0.
  \end{eqnarray}
  If one uses the identity
$(p \cdot \Ren{a}_p) \, \frac{\partial}{\partial \tau}
    + p \cdot \Ren{\nabla} = p^\mu\,\partial_\mu$,
Eq.(\ref{eq:1-061}) is found to have the following form
  \begin{eqnarray}
    \label{eq:1-064}
    \partial_\mu T^{\mu\nu}_{\mathrm{1st}} = 0,\quad
    \partial_\mu N^{\mu}_{\mathrm{1st}} = 0.
  \end{eqnarray}
  with
  \begin{eqnarray}
    \label{eq:1-065}
    T^{\mu\nu}_{\mathrm{1st}}
    &\equiv& \sum_{p} \, \frac{1}{p^0} \, p^\mu \, \varphi_{0p}^{\nu}
    \Big\{ f^{\mathrm{eq}}_p - \big[ A^{-1} \, \bar{Q}_0 \, F \big]_p \Big\}, \\
    N^{\mu}_{\mathrm{1st}}
    &\equiv& m^{-1}\,\sum_{p} \, \frac{1}{p^0} \, p^\mu \, \varphi_{0p}^{4}
    \Big\{ f^{\mathrm{eq}}_p - \big[ A^{-1} \, \bar{Q}_0 \, F \big]_p \Big\}.
  \end{eqnarray}

\subsection{Example: Landau-Lifshitz frame}

In this subsection, we present the hydrodynamic equation  
 derived by the RG method.

If we take the most natural choice
for the macroscopic vector as $\Ren{a}^\mu(x)=u^{\mu}(x)$, 
the resultant energy-momentum tensor and particle current turn out to be\cite{env009,Tsumura:2011cj}
  \begin{eqnarray}
    \label{eq:2-068}
    T^{\mu\nu}_{\mathrm{1st}} &=& e\,u^\mu\,u^\nu - (p - \zeta \, \nabla\cdot u)\,\Delta^{\mu\nu}
    + 2 \, \eta \, \Delta^{\mu\nu\rho\sigma} \, \nabla_\rho u_\sigma,\\
    \label{eq:2-069}
    N^\mu_{\mathrm{1st}} &=& n\,u^\mu + \lambda \,
    \frac{1}{\hat{h}^2} \, \nabla^\mu\frac{\mu}{T},
  \end{eqnarray}
respectively,
with $\Delta^{\mu\nu} \equiv g^{\mu\nu} - u^\mu\,u^\nu$,
$\nabla^\mu \equiv \Delta^{\mu\nu}\,\partial_\nu$,
and
$\Delta^{\mu\nu\rho\sigma} \equiv 
1/2 \cdot (\Delta^{\mu\rho}\Delta^{\nu\sigma} + \Delta^{\mu\sigma}\Delta^{\nu\rho}
- 2/3 \cdot \Delta^{\mu\nu}\Delta^{\rho\sigma})$.
Here,
$n$, $p$, and $\hat{h}$ denote
the particle-number density, pressure, and reduced enthalpy per particle,
respectively.
 It is clear that these formulae  completely agree with those 
 proposed by Landau and Lifshitz \cite{hen002}.
  Indeed,
  the respective dissipative parts
  $\delta T^{\mu\nu}$ and $\delta N^{\mu}$
  in Eqs. (\ref{eq:2-068}) and (\ref{eq:2-069}) meet Landau and Lifshitz's constraints
\beq
\delta e \equiv u_\mu  \delta T^{\mu\nu} u_\nu = 0,\quad
\delta n \equiv u_\mu  \delta N^\mu = 0,\quad
Q_\mu \equiv \Delta_{\mu\nu} \delta T^{\nu\rho} u_\rho = 0.
\eeq
Thus we have derived the dissipative hydrodynamics in the 
Landau-Lifshitz (energy) frame in the RG method.

Moreover, since our theory is in the level of  statistical mechanics,
we have the microscopic expressions for 
the transport coefficients appearing in the hydrodynamic tensor(\ref{eq:2-068}) and current 
(\ref{eq:2-069}), as follows:
\beq
\zeta = - \frac{1}{T} \langle  \tilde{\Pi},L^{-1}\tilde{\Pi} \rangle,\quad
\lambda = \frac{1}{3 T^2} \langle  \tilde{J}^\mu, L^{-1}\tilde{J}_\mu \rangle,
\quad
\eta = - \frac{1}{10 T}  \langle \tilde{\pi}^{\mu\nu}, L^{-1}\tilde{\pi}_{\mu\nu} \rangle.
\label{eq:trans-coef-1st}
\eeq
  Here, we have introduced the following microscopic thermal forces
  $(\tilde{\Pi}_p,\,\tilde{J}^\mu_p,\,\tilde{\pi}^{\mu\nu}_p) \equiv
 (\Pi_p,\,J^\mu_p,\,\pi^{\mu\nu}_p)/{(p\cdot u)}$, where
\beq
\Pi_p \equiv \Big( \frac{4}{3} - \gamma \Big) \, (p \cdot u)^2
    + \Big( (\gamma - 1) \, T \, \hat{h} - \gamma \, T \Big) \, (p \cdot u)
    - \frac{1}{3} \, m^2,
\eeq
$
J^\mu_p \equiv - ( (p \cdot u) - T \, \hat{h} ) \, \Delta^{\mu\nu} \, p_\nu$
and
$\pi^{\mu\nu}_p \equiv \Delta^{\mu\nu\rho\sigma} \, p_\rho \,  p_\sigma$.
Here,
$\gamma$ denotes
the ratio of the heat capacities.

It is noteworthy that
the transport coefficients can be
rewritten in the Green-Kubo formula \cite{text}.
With the use of  the ``time-dependent thermal force'' defined by
$\tilde{\Pi}_p(s)\equiv  \sum_q \left[ e^{sL} \right]_{pq}\tilde{\Pi}_q$
 and so on,  the relaxation functions are given by the time-correlators 
\beq
    \label{eq:1-037}
R_\zeta(s) &\equiv& \frac{1}{T} \, \langle \, \tilde{\Pi}(0)\,,\,\tilde{\Pi}(s) \,
    \rangle,
\eeq
and so on for $R_\lambda(s)$ and $R_\eta(s)$ with obvious modifications to $R_{\zeta}(s)$.
Then the transport coefficients given in Eq.(\ref{eq:trans-coef-1st}) are rewritten as
follows,
  \begin{eqnarray}
    \zeta = \int_0^\infty\!\!ds\,\,\,R_\zeta(s),\quad
    \lambda = \int_0^\infty\!\!ds\,\,\,R_\lambda(s),\quad
    \eta = \int_0^\infty\!\!ds\,\,\,R_\eta(s).
  \end{eqnarray}

We remark that the instability problem of the local equilibrium state is 
analyzed in \citen{Tsumura:2011cj} and \citen{env010}.

\section{Second-order equations and moment method}

In the derivation of the first-order hydrodynamic equations like Landau-Lifshits
equation, we utilized the zero modes of the linearized collision operator,
which form the invariant manifold on which hydrodynamics is defined;
the would-be constant zero modes acquire the time-dependence on the manifold by the
RG equation. Our formalism can be extended to include excited modes as additional
components of the invariant/attractive manifold\cite{next002}. 
The out-come is nothing but the extended
thermodynamics or Israel-Stewart type equation, with new microscopic expressions
of the relaxation times and lengths. Furthermore,
we find the proper ansatz for the distribution function to be used in the
moment method\cite{next002}. 
For the shortage of space, we here only present the results,
leaving the detailed derivation in a separate paper\cite{next002}.
We emphasize that our theory gives an explicit construction of 
the invariant manifold corresponding to thirteen moments, 
which has been long sought for\cite{meso,Gorban}.

\subsection{
A brief review of  Grad's thirteen-moment method 
and Grad-Muller equation: non-relativistic case
}
\label{sec:002-2}
In Grad's thirteen-moment method\cite{grad,mic001},
the one-particle distribution function $f_{\Vec{v}}(t,\,\Vec{x})$ is 
expressed in terms of the equilibrium distribution function $f^{\mathrm{eq}}_{\Vec{v}}(t,\,\Vec{x})$ as
$f_{\Vec{v}}(t,\,\Vec{x}) = f^{\mathrm{eq}}_{\Vec{v}}(t,\,\Vec{x})\,\Big(1 +
  \Phi_{\Vec{v}}(t,\,\Vec{x})\Big)$,
where $\Phi_{\Vec{v}}(t,\,\Vec{x})$ is assumed to have the form
\beq
\Phi_{\Vec{v}}(t,\,\Vec{x}) = \Phi^{\mathrm{G}}_{\Vec{v}}(t,\,\Vec{x})
  \equiv \hat{\pi}^{ij}_{\Vec{v}}(t,\,\Vec{x}) \, \pi^{ij}(t,\,\Vec{x})
  + \hat{J}^{i}_{\Vec{v}}(t,\,\Vec{x}) \, J^{i}(t,\,\Vec{x}).
\eeq
Here,
$\hat{\pi}^{ij}_{\Vec{v}}(t,\,\Vec{x})$ and $\hat{J}^i_{\Vec{v}}(t,\,\Vec{x})$
are defined by
$\hat{\pi}^{ij}_{\Vec{v}}(t,\,\Vec{x}) \equiv
  m \,(\delta v^i(t,\,\Vec{x}) \, \delta
  v^j(t,\,\Vec{x}) - \frac{1}{3} \, \delta^{ij} \, |\Vec{\delta v}(t,\,\Vec{x})|^2)$
and
$\hat{J}^i_{\Vec{v}}(t,\,\Vec{x}) \equiv
  (\frac{m}{2}\,
  |\Vec{\delta v}(t,\,\Vec{x})|^2 - \frac{5}{2}\,T(t,\,\Vec{x}))\,\delta v^i(t,\,\Vec{x})$,
with
$\Vec{\delta v}(t,\,\Vec{x}) \equiv \Vec{v} - \Vec{u}(t,\,\Vec{x})$ being the peculiar 
velocity.

Then the evolution equation of the thirteen coefficients
are determined by the equations all of which are derived from 
the Boltzmann equation
with use of 
the linearized collision operator given by
$L_{\Vec{v}\Vec{k}} \equiv (f^{\mathrm{eq}}_{\Vec{v}})^{-1}\,
   \frac{\partial}{\partial f_{\Vec{k}}}C[f]_{\Vec{v}}\Bigg|_{f=f^{\mathrm{eq}}} \,
   f^{\mathrm{eq}}_{\Vec{k}}$.
Thus the Grad-Muller equation
is obtained
as a closed system of the equations governing 
$T$, $n$, $u^i$, $\pi^{ij}$, and $J^i$, which includes
\begin{eqnarray}
  m\,n\,\frac{\partial}{\partial t}u^i
  +m\,n\,\Vec{u}\cdot\Vec{\nabla}u^i
  &=& - \nabla^j(p\,\delta^{ji} - 2\,\eta^{\mathrm{G}}\,\pi^{ji}),\\
  \pi^{ij} + \tau^{\mathrm{G}}_{\pi} \, 
\Big(\frac{\partial}{\partial t} + \Vec{u}\cdot\Vec{\nabla}\Big)\pi^{ij} &=&
\bar{X}^{ij}_{\pi}
  + (\mathrm{other \, terms}),
\end{eqnarray}
with $p$ being the pressure, 
where we have defined the thermodynamic force given by
$\bar{X}^{ij}_\pi \equiv 1/2\cdot(\nabla^i u^j + \nabla^j u^i - 2/3 \cdot \delta^{ij} \, \Vec{\nabla} \cdot \Vec{u})$.

In the Grad moment theory,
the transport coefficients and relaxation times are given in terms of 
the inner product defined by
${\langle\, \psi\,,\, \chi\,\rangle}_{\mathrm{eq}} \equiv \sum_{\Vec{v}} \, f^{\mathrm{eq}}_{\Vec{v}}
   \, \psi_{\Vec{v}} \, \chi_{\Vec{v}}$.
For example,
\begin{eqnarray}
  \label{eq:moment-12}
  \eta^{\mathrm{G}} =
 - \frac{1}{10\,T}\,\frac{{\langle\, \hat{\pi}^{ij}\,,\,\hat{\pi}^{ij} \,\rangle}_{\mathrm{eq}}
  \, {\langle\, \hat{\pi}^{kl}\,,\,\hat{\pi}^{kl} \,\rangle}_{\mathrm{eq}}}{
  {\langle\, \hat{\pi}^{mn}\,,\,L \, \hat{\pi}^{mn} \,\rangle}_{\mathrm{eq}}},\quad
  \tau^{\mathrm{G}}_{\pi} = -\frac{{\langle\, \hat{\pi}^{ij}\,,\,\hat{\pi}^{ij} \,\rangle}_{\mathrm{eq}}}
  {{\langle\, \hat{\pi}^{kl}\,,\,L\,\hat{\pi}^{kl} \,\rangle}_{\mathrm{eq}}}.
  \end{eqnarray}
These formulae are different from 
 those given in the Chapman-Enskog expansion method; for instance,
 $\eta^{\mathrm{CE}} = - {1}/{10\,T}\,\cdot
     {\langle\, \hat{\pi}^{ij}\,,\,L^{-1} \, \hat{\pi}^{ij} \,\rangle}_{\mathrm{eq}}$,
which clearly shows that
$\eta^{\mathrm{CE}} \ne \eta^{\mathrm{G}}$.

  \subsection{
The mesoscopic dynamics from the RG method: nonrelativistic case}

We apply the RG method to derive 
the extended thermodynamics or mesoscopic dynamics\cite{meso,Gorban} 
from the nonrelativistic Boltzmann equation, which is equivalent to 
construct the invariant manifold\cite{env006} in the space of the distribution function.
It is nice that we can 
read off the proper ansatz for the deviation function $\Phi_{\Vec{v}}$ from 
the distribution function constructed in our method.
Leaving the detailed derivation of the formulae 
to a separate paper\cite{next002}, we here give the results:
Our deviation function $\Phi_{\Vec{v}}$ and the relaxation equations read
  \begin{eqnarray}
    \label{eq:moment-19}
    \Phi_{\Vec{v}} =
    \Phi^{\mathrm{TK}}_{\Vec{v}}
    \equiv \frac{1}{T}\,\big[ L^{-1} \, \hat{\pi}^{ij} \big]_{\Vec{v}} \, \pi^{ij}
    + \frac{1}{T}\,\big[ L^{-1} \, \hat{J}^{i} \big]_{\Vec{v}} \, J^{i},
  \end{eqnarray}
  and  
  \begin{eqnarray}
     \label{eq:moment-20}
     \sum_{\Vec{v}} \, \frac{1}{T}\,\big[ L^{-1} \, \hat{\pi}^{ij} \big]_{\Vec{v}} \,
     \Bigg[ \frac{\partial}{\partial t} + \Vec{v}\cdot\Vec{\nabla} \Bigg] f_{\Vec{v}}
     &=& \sum_{\Vec{v}} \, \sum_{\Vec{k}} \, f^{\mathrm{eq}}_{\Vec{v}} \,
     \frac{1}{T}\,\big[ L^{-1} \, \hat{\pi}^{ij} \big]_{\Vec{v}} \, L_{\Vec{v}\Vec{k}}
     \, \Phi_{\Vec{k}},\\
     \label{eq:moment-21}
     \sum_{\Vec{v}} \, \frac{1}{T}\,\big[ L^{-1} \, \hat{J}^{i} \big]_{\Vec{v}} \,
     \Bigg[ \frac{\partial}{\partial t} + \Vec{v}\cdot\Vec{\nabla} \Bigg] f_{\Vec{v}}
     &=& \sum_{\Vec{v}} \, \sum_{\Vec{k}} \, f^{\mathrm{eq}}_{\Vec{v}} \,
     \frac{1}{T}\,\big[ L^{-1} \, \hat{J}^{i} \big]_{\Vec{v}} \, L_{\Vec{v}\Vec{k}}
     \, \Phi_{\Vec{k}},
  \end{eqnarray}
  respectively.
  The corresponding  microscopic expressions of the transport coefficients are given by
  \begin{eqnarray}
     \eta^{\mathrm{TK}} = - \frac{1}{10\,T}\,
     {\langle\, \hat{\pi}^{ij}\,,\,L^{-1} \, \hat{\pi}^{ij} \,\rangle}_{\mathrm{eq}},\, \quad
     \lambda^{\mathrm{TK}} =
 - \frac{1}{3\,T^2}\,{\langle\, \hat{J}^{i}\,,\,L^{-1} \, \hat{J}^{i} \,\rangle}_{\mathrm{eq}},
  \end{eqnarray}
  which perfectly agree with those by Chapman-Enskog method,
  $\eta^{\mathrm{CE}}$ and $\lambda^{\mathrm{CE}}$.
  Furthermore, we have the microscopic representation of the relaxation times given by
  \begin{eqnarray}
    \tau^{\mathrm{TK}}_{\pi} =
 -\frac{{\langle\, \hat{\pi}^{ij}\,,\,L^{-2} \, \hat{\pi}^{ij} \,\rangle}_{\mathrm{eq}}}
    {{\langle\, \hat{\pi}^{kl}\,,\,L^{-1}\,\hat{\pi}^{kl} \,\rangle}_{\mathrm{eq}}},\, \quad
    \tau^{\mathrm{TK}}_{J} = -\frac{{\langle\, \hat{J}^{i}\,,\,L^{-2}\,\hat{J}^{i} \,\rangle}_{\mathrm{eq}}}
    {{\langle\, \hat{J}^{k}\,,\,L^{-1}\,\hat{J}^{k} \,\rangle}_{\mathrm{eq}}},
  \end{eqnarray}
which are new and different from those in the naive Grad moment method.

\subsection{The mesoscopic dynamics from the RG method: relativistic case}

In the relativistic case,
if we express the distribution function by
$f_p(x) = f^{\mathrm{eq}}_p(x)\,\big(1 + \Phi_p(x)\big)$,
our RG method gives the following expression 
of $\Phi_p(x)$,
\begin{eqnarray}
  \Phi_p = -\frac{1}{T}\,\sum_q
  \, L^{-1}_{pq}\,
  (\tilde{\Pi}_q \, \Pi  + \tilde{J}^\mu_q \, J_\mu +
  \tilde{\pi}^{\mu\nu}_q \, \pi_{\mu\nu}),
\end{eqnarray}
where $\tilde{\Pi}_p$, $\tilde{J}^\mu_p$,
and $\tilde{\pi}^{\mu\nu}_p$ are the microscopic thermal forces,
This form is different from those used by
Israel-Stewart\cite{mic004} and Denicol et al.\cite{Denicol:2010xn}
within the  moment method.

The relaxation equations derived in our RG method read
\begin{eqnarray}
  \Pi &=& -\nabla\cdot u
- \tau_\Pi \,  D\Pi +
\textrm{other terms involving relaxation lengths},\nonumber\\
  J^\mu &=&
\frac{1}{\hat{h}^2} \, \nabla^\mu \frac{\mu}{T}
  - \tau_J \, \Delta^{\mu a} \, DJ_a+
\textrm{other terms involving relaxation lengths},
  \nonumber\\
  \pi^{\mu\nu} &=&
\Delta^{\mu\nu\rho\sigma} \, \nabla_\rho u_\sigma
- 
\tau_\pi \, \Delta^{\mu\nu ab} \, D\pi_{ab} 
 \nonumber\\
  &&{}+ \Big(\kappa^{(0)}_{\pi\pi}\,\Delta^{\mu\nu\rho\sigma}\,\nabla\cdot u
  + \kappa^{(1)}_{\pi\pi} \, \Delta^{\mu\nu ac}\,
  \Delta_c^{\,\,\,b\rho\sigma} \, \Delta_{abde}\,\nabla^d u^e
  + \kappa^{(2)}_{\pi\pi}\,\Delta^{\mu\nu ac}\,
  \Delta_c^{\,\,\,b\rho\sigma} \, \omega_{ab}\Big) \, \pi_{\rho\sigma}\nonumber \\
 &&{}+\textrm{other terms involving relaxation lengths},
\end{eqnarray}
with the volticity 
$\omega^{\mu\nu} \equiv \frac{1}{2} \, (\nabla^\mu u^\nu - \nabla^\nu u^\mu)$
and $D \equiv u^\mu \, \partial_\mu$.
The energy-momentum tensor and particle current read
$T^{\mu\nu}_{\mathrm{2nd}} = e\,u^\mu\,u^\nu - (p + \zeta \, \Pi)\,\Delta^{\mu\nu} + 2 \, \eta \, \pi^{\mu\nu}$
and $N^\mu_{\mathrm{2nd}} = n\,u^\mu + \lambda \,J^\mu$, respectively.

The RG method gives microscopic expressions for the resulting times
 $\tau_\Pi$, $\tau_J$, and $\tau_\pi$ as follows
\begin{eqnarray}
  \tau_\Pi &\equiv&
  - \frac{
    \langle \, \tilde{\Pi}\,,\,L^{-2}\,\tilde{\Pi} \, \rangle
  }{
    \langle \, \tilde{\Pi}\,,\,L^{-1}\,\tilde{\Pi} \, \rangle
  }
  =\frac{
    \int_0^\infty\!\!\mathrm{d}s\,\,\,s\,R_\zeta(s)
  }{
    \int_0^\infty\!\!\mathrm{d}s\,\,\,R_\zeta(s)
  }  ,
\eeq
and so on for 
$\tau_J$ and $\tau_{\pi}$ with obvious modifications.
It is noteworthy that
the relaxation times $\tau_\Pi$, $\tau_J$, and $\tau_\pi$
are represented in terms of the relaxation functions $R_\zeta(s)$, $R_\lambda(s)$ and
$R_\eta(s)$:  These formulae have a clear physical meaning of
the relaxation times as the  correlated times in the relaxation function.
We note that
it is for the first time that
such representation and the interpretation of
the relaxation times are given for the extended thermodynamics,
 as far as we are aware of.
We also note that shear viscosity derived in our method do 
coincide with that in the Chapman-Enskog method as shown before.

\section{Brief summary}
We have reported our attempts to derive first-order and second-order
relativistic hydrodynamic equations from  relativistic Boltzmann equation which 
has a manifest Lorentz invariance and does not show any pathological 
behavior such as the instability and acausality seen in existing hydrodynamic equations.
We have given the novel extended thermodynamics both in nonrelativistic and
relativistic cases through the explicit construction of attractive manifold containing
the relaxation process from Boltzmann equation. 

\section*{Acknowledgements}
  We are grateful to K. Ohnishi for his collaboration in the early stage of this work.
  T.K. was partially supported by a
  Grant-in-Aid for Scientific Research from the Ministry of Education,
  Culture, Sports, Science and Technology (MEXT) of Japan
  (Nos. 20540265 and 23340067),
  by the Yukawa International Program for Quark-Hadron Sciences, and by a
  Grant-in-Aid for the global COE program
  ``The Next Generation of Physics, Spun from Universality and Emergence'' from MEXT.

%

\end{document}